\documentclass[amsmath,amssymb,nofootinbib,twocolumn]{revtex4}

\usepackage{latexsym,comment,verbatim}
\usepackage{amssymb}
\usepackage{amsfonts}
\usepackage{amsmath,color}
\usepackage[dvips]{graphicx}
\usepackage{bm}

\def\mb#1{\mathbf{#1}}

\def\ber{\begin{eqnarray}}
\def\eer{\end{eqnarray}}
\def\beq{\begin{equation}}
\def\eeq{\end{equation}}

\def\ed{\end{document}}

\def\dT#1{\frac{\mathrm{d} #1}{\mathrm{d}T}}

\def\sT{\sin \left(\omega T \right)}
\def\cT{\cos \left(\omega T \right)}
\def\Di#1#2{\frac{\mathrm{D} #1}{\mathrm{d}#2}}

\begin{document}

\title{Gravito-magnetic resonance in the field of a gravitational wave}

\author{Matteo Luca Ruggiero}
\email{matteo.ruggiero@polito.it}
\affiliation{Politecnico di Torino, Corso Duca degli Abruzzi 24, Torino, Italy}
\author{Antonello Ortolan}
\affiliation{INFN - National Laboratories of Legnaro, Viale dell'Universit\`a 2, I-35020 Legnaro (PD), Italy}


\begin{abstract}
Using the construction of the Fermi frame, the field of a gravitational wave can be described in terms of gravito-electromagnetic fields that are transverse  to the propagation direction and orthogonal to each other. In particular, the gravito-magnetic field acts on spinning particles and we show that, due to the action of the gravitational wave field, a new phenomenon, that we call gravito-magnetic resonance, may appear. We give both a classical and quantum description of this  phenomenon and suggest that it can be used as the basis for new type of gravitational wave detectors.    Our results  highlight the effectiveness  of  collective spin excitations, e.g.  spin waves in magnetized materials,  in detecting high frequency gravitational waves.  Here we suggest that, when gravitational waves induce a precession of the electron spin,   power is released in the ferromagnetic resonant mode endowed with  quadrupole symmetry of a magnetized sphere. This offers a possible path to the detection of the gravito-magnetic effects of a gravitational wave. 
\end{abstract}

\maketitle

\section{Introduction}\label{sec:intro}

The recent detection of gravitational waves \cite{PhysRevLett.116.061102,PhysRevLett.119.161101} by means of the large interferometers, confirms once again the predictions of General Relativity and the role of Einstein's theory as the best model of gravitational interactions, even though large scale astronomical observations keep challenging the Einsteinian paradigm  with the problems of dark matter and dark energy \cite{2014arXiv1409.7871W,universe2040023}. The measurement process needs to be carefully analysed in General Relativity, since it becomes meaningful only when the observer and the object of observation are  clearly identified \cite{de2010classical}. This is of uttermost importance when dealing with gravitational waves, since their effects on terrestrial detectors are very small: misconceptions may arise \cite{Faraoni:2007gr} or subtleties  need to be properly taken into account \cite{Finn:2008np}.  One possible approach is to use a Fermi coordinates system, defined in the vicinity of the world-line of an observer arbitrarily moving in spacetime. Even though the interaction between gravitational waves and detectors is usually studied using the so-called transverse and traceless coordinates,  other approaches are viable \cite{Rakhmanov_2014}. In particular, Fermi coordinates have a direct operational meaning, since they are the coordinates an observer would use to perform space and time measurements; indeed, using these coordinates the metric tensor contains (up to the required approximation level) only quantities that are invariant under coordinate transformations internal to the reference frame. 

Actually, beyond the aforementioned difficulties for Einstein's theory deriving from cosmological observations, it is a matter of fact that, up today,  the interplay between General Relativity and Quantum Mechanics is not clear.  How could we reconcile the Standard Model of particle physics with the geometrical description of the gravitational interaction? There are no conclusive answers from the theoretical point of view and, also, there are no observations whose explanations require both General Relativity and Quantum Mechanics. In this context, spinning point-like particles have an important role. In fact, while intrinsic spin is a fundamentally quantum property, in Einstein's theory spin is present only at classical level deriving from the rotation of finite-size bodies \cite{Fadeev:2020gjk}. So, General Relativity, as is, does not explicitly describe the interaction of spacetime with spinning point-like particles; in particular, a relevant question is whether these particles undergo gravito-magnetic effects \cite{Ruggiero:2002hz}.  As suggested by \citet{Mashhoon:2003ax}, this question is related to the inertia of intrinsic spin; more generally, the spin-gravity coupling is related to the spin-rotation coupling, on the basis of Einstein's principle of equivalence.  In a spacetime that is almost locally flat, around the world-line of an observer, this interaction can be described in terms of coupling between the gravito-magnetic field and the intrinsic spin, and it can also  be obtained  from suitable limits of a Dirac-type equation \cite{hehl1990inertial}. Any experiments aimed at testing effects of gravity on spinning particles is indeed a test of the equivalence principle in a new regime \cite{Tasson:2012nx}. If we go beyond General Relativity,  the theoretical possibilities increase; for instance, in the Einstein-Cartan theory the role of spin is to generate (non propagating) torsion, while mass and energy determine curvature \cite{Hehl:1976kj,Ruggiero:2003tw}.  More generally, spin-gravity interaction is peculiar in extended theories of gravity \cite{Capozziello:2011et} and  there are many experiments of fundamental physics \cite{Safronova:2017xyt}, in which various theoretical possibilities are investigated.

It is then manifest that investigating the interaction between gravity and spin could shed new light on the interplay between our best model of gravitational interaction, General Relativity, and the Standard Model of particle physics, which is based on Quantum Mechanics. In a previous paper \cite{Ruggiero_2020} we showed that the effects of a plane gravitational wave can be described, in a Fermi coordinate system, in terms of a gravito-electromagnetic analogy. Namely, we illustrated that the wave field is equivalent to the action of a gravito-electric and a gravito-magnetic field, that are transverse to the propagation direction and orthogonal to each other. Hence the interaction with detectors is operationally defined by a (tidal) gravito-electromagnetic Lorentz force. In particular, all current detectors, such LIGO and VIRGO, or foreseeable ones, like LISA, are indeed aimed at revealing the interaction of a system of masses with the electric-like component of the field. However, due to the magnetic-like component, there is an interaction with moving masses and spinning particles \cite{biniortolan2017}. This is not a surprise, since a gravitational wave transports angular momentum.
There have been recent proposals to use spinning particles as a probe for gravitational waves \cite{Ito:2020wxi}, however they are based on the tidal gravito-electric effect and, hence, quite similar to  the detection methods of the interferometers. 

Here, we propose a new effect: focusing on the interaction of a gravitational wave with a spinning particle, we show that, in analogy with what happens in electromagnetism, a gravito-magnetic resonance phenomenon may appear and we suggest that this effect can be exploited, in principle, to design new types of detectors of gravitational waves.

The paper is organised as follows: in Section \ref{sec:gmf}  we introduce the gravito-electromagnetic approach in the Fermi frame, while in Section \ref{sec:gmr} we introduce the phenomenon of gravito-magnetic resonance for spinning particles in the field of a gravitational wave. Conclusions are eventually in Section \ref{sec:dconc}.

\section{Gravito-electromagnetic fields in the Fermi frame} \label{sec:gmf}

Fermi coordinates in the vicinity of an observer's world-line can be defined as follows  \cite{MTW,marzlin}. In the background spacetime describing the gravitational field, we consider a set of coordinates\footnote{Greek indices refer to spacetime coordinates, and assume the values $0,1,2,3$, while Latin indices refer to spatial coordinates and assume the values $1,2,3$, usually corresponding to the coordinates $x,y,z$.} $x^{\mu}$; accordingly, the world-line $x^{\mu}(\tau)$ of a reference observer as function of the proper time $\tau$ is determined by the following equation $\Di{x^{\mu}}{\tau}=\ddot x^{\mu}+\Gamma^{\mu}_{ \ \nu\sigma}\dot x ^{\nu} \dot x ^{\sigma} = a^{\mu}$, where $\mathrm D$ stands for the covariant derivative along the world-line, a dot means derivative with respect to $\tau$ and $a^{\mu}$ is the four-acceleration. In the tangent space along the world-line $x^{\mu}(\tau)$ we define the orthonormal tetrad of the observer $e^{\mu}_{(\alpha)}(\tau)$ such that $e^{\mu}_{(0)}(\tau)$ is the unit vector tangent to his world-line and $e^{\mu}_{(i)}(\tau)$ (for i=1,2,3) are the spatial vectors  orthogonal to each other and, also, orthogonal to $e^{\mu}_{(0)}(\tau)$. The equation of motion of the tetrad is $\Di{e^{\mu}_{(\alpha)}}{\tau}=-\Omega^{\mu\nu} e_{\nu (\alpha)}$ where $\Omega^{\mu\nu}=a^{\mu} \dot x ^{\nu} - a^{\nu}\dot x^{\mu}+\dot x_{\alpha}\Omega_{\beta}\epsilon^{\alpha\beta\mu\nu}$. In the latter equation $\Omega^{\alpha}$ is the four-rotation of the tetrad. In particular, we notice that for a geodesic ($a^{\mu}=0$) and non rotating ($\Omega^{\alpha}=0$) tetrad we have  $\Omega^{\mu\nu}=0$: consequently, in this case the tetrad is parallel transported. If $\Omega=0$ and $a^{\mu} \neq 0$, the tetrad is Fermi-Walker transported; indeed, Fermi-Walker transport enables to define the natural \textit{non rotating} moving frame for an accelerated observer \cite{MTW}. Fermi coordinates are defined within a cylindrical spacetime region of radius $\mathcal R$, in the vicinity of the reference world-line, where  $\mathcal R$ is the spacetime radius of curvature: the observer along the congruence measures time intervals according to the proper time, so the time coordinate is defined by $T=\tau$; the spatial coordinates $X,Y,Z$ are defined by space-like geodesics, with unit tangent vectors $n^{\mu}$, whose components, with respect to the orthonormal tetrad are $n^{(i)}=n_{(i)}=n_{\mu}e^{\mu}_{(i)}(\tau)$, and $n^{(0)}=0$. The reference observer's frame equipped with Fermi coordinates is the \textit{Fermi frame}. Fermi coordinates in the vicinity of the world-line of an observer in accelerated motion  with rotating tetrads  were studied in  \cite{Ni:1978di,Li:1979bz,marzlin}.  It is possible to show that  the spacetime element in Fermi coordinates in the vicinity of the observer's world-line can be recast  in terms of the gravito-electromagntic  potentials $(\Phi, \mb A)$  \cite{Mashhoon:2003ax,Ruggiero_2020}
\begin{equation} 
ds^2=-\left(1-2\frac{\Phi}{c^2}\right)c^{2}dT^2-\frac{4}{c}({\mb 
A}\cdot d{\mb
X})dt+\delta_{ij}dX^idX^j \, , \label{eq:mmetric2}
\end{equation}
 and this peculiarity allows us to apply the corresponding formalism to spinning particles\footnote{Here and henceforth $\mb X$ is the position vector in the Fermi frame.}.  The gravito-electromagntic  potentials depend both on the inertial features of the Fermi frame, through  $\mb a$ and $\bm \Omega$,  i.e. the projection of the observer's acceleration and tetrad rotation onto the local frame, respectively, and on the spacetime curvature, through the Riemann curvature tensor.
Here we are interested in the gravito-magnetic effects; the gravito-magnetic potential  is $ A_{i}(T, {\mb X})=A_{i}^{I}(\mb X)+A^{C}_{i}(T ,{\mb X})$, where the  \textit{inertial} contribution is  $  A_{i}^{I}(\mb X)=-\left(\frac{\mb \Omega c}{2} \wedge \mb X\right)_{i}$ and the  \textit{curvature} contribution is $ A^{C}_{i}(T ,{\mb X})=\frac{1}{3}R_{0jik}(T )X^jX^k$. Accordingly, it is possible to define the gravito-magnetic field $ \mb B=\mb B^I+\mb B^C$ where
\beq
 B_{i}^{I} = - \Omega_{i} c, \quad B^{C}_i(T ,{\mb R})=-\frac{c^{2}}{2}\epsilon_{ijk}R^{jk}_{\;\;\;\; 0l}(T )X^l. \label{eq:defBIBG}
\eeq

\section{Gravito-magnetic resonance} \label{sec:gmr}

Exploiting the gravito-electromagnetic analogy, we may say that a test spinning particle with mass $m$ and spin $\mb S$ has a gravito-magnetic charge $q_{B}=-2m$ and, as a consequence, it possesses a \textit{gravito-magnetic dipole moment} $\displaystyle \bm \mu_{g}=-\frac{\mb S}{c}$ \cite{Mashhoon:2003ax}. Hence, in an external gravito-magnetic field $\mb B$, its evolution equation is
\beq
\dT{\mb S}= \bm \mu_{g} \times \mb B = -\frac{1}{c} \mb S\times \mb B = \frac{1}{c} \mb B \times \mb S. \label{eq:spinevol0}
\eeq 

\begin{figure}[t]
\begin{center}
\includegraphics[scale=.50]{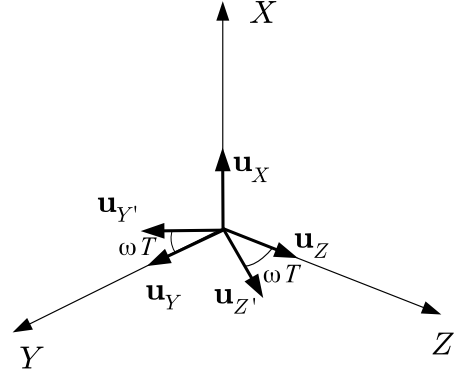}
\caption{The Fermi frame is equipped with spatial coordinates $X,Y,Z$,  with unit vectors $\{ \mb u_{X}, \mb u_{Y}, \mb u_{Z} \}$; the unit vectors $\{\mb u'_{Y}, \mb u'_{Z} \}$ are rotating with frequency $\omega$.} \label{fig:assi}
\end{center}
\end{figure}

Now, let us consider a spinning particle interacting with the wave gravito-magnetic field. In the Fermi frame, we consider the coordinates $T,X,Y,Z$ defined as above, with a set of unit vectors $\{ \mb u_{X}, \mb u_{Y}, \mb u_{Z} \}$, and we suppose that  the plane gravitational wave propagates along the $X$ axis (see Figure \ref{fig:assi}). As discussed in \cite{Ruggiero_2020},  the curvature part of the gravito-magnetic field (\ref{eq:defBIBG}) has the following components:
\begin{widetext}
\beq
B^{C}_{X}  = 0, \quad B^{}_{Y}  = -\frac{\omega^{2}}{2}\left[-A^{\times} \cT Y+A^{+} \sT Z \right], \quad B^{}_{Z}  = -\frac{\omega^{2}}{2}\left[A^{+}\sT Y+A^{\times}\cT Z \right]. \label{eq:defBxyz1}
\eeq
\end{widetext}

{In the above definitions $A^{+}$  and $A^{\times}$ are the amplitude of the wave in the two polarisation states, while $\omega$ is its frequency. In order to evaluate the effects, we consider a circularly polarized wave, so that $A^{+}=A^{\times}=A$ and  the  field (\ref{eq:defBxyz1}) becomes
\begin{widetext}
\beq
B^{C}_{X}  = 0, \quad B^{}_{Y}  = -\frac{A\omega^{2}}{2}\left[- \cT Y+ \sT Z \right], \quad B^{}_{Z}  = -\frac{A\omega^{2}}{2}\left[\sT Y+\cT Z \right]. \label{eq:defBxyz12}
\eeq
\end{widetext}
If we consider a frame  clockwise rotating  in the $YZ$ plane with the wave frequency $\omega$, its basis vectors are given by $\mb u_{X'}=\mb u_{X}$, $\mb u_{Y'} (T)= \cT \mb u_{Y}-\sT \mb u_{Z}$ and $\mb u_{Z'} (T)= \sT \mb u_{Y}+ \cT \mb u_{Z}$ (see Figure \ref{fig:assi}). The above field (\ref{eq:defBxyz12}) can be written in the form
\beq
 \mb B^{C}(T)=\frac{A\omega^{2}}{2}\left[Y\mb u_{Y'}(T)-Z\mb u_{Z'}(T)\right]
\eeq
The magnitude of this field is $\displaystyle B_{C}=\frac{A\omega^{2}}{2} \sqrt{Y^{2}+Z^{2}}=\frac{A\omega^{2}}{2} L$, where $L$ is the distance from the origin of the frame; the gravito-magnetic field $ \mb B^{C}$  is clockwise rotating  with the wave frequency $\omega$, but  is a \textit{static field} in the considered rotating frame.}

As we have seen, in the Fermi frame  the total gravito-magnetic field is $\mb B=\mb B^{I}+\mb B^{C}$, where $\mb B^{I} = -\mb \Omega c$ and it is simply proportional to the rotation rate $\mb \Omega$ of the frame. We suppose that the latter field is static and that  the frame rotates along the direction of propagation of the wave; hence, we may write  $\mb B^{I}=-B^{I} \mb u_{X}$ where $B^{I}=\Omega c$.  Accordingly, the evolution of a spinning test particle is determined by  Eq. (\ref{eq:spinevol0}), which in this case becomes 
\beq
\dT{\mb S}=  \frac{1}{c} \left[\mb B^{C}(T)+ \mb B^{I} \right] \times \mb S. \label{eq:spinevol1}
\eeq 

{If we consider the frame co-rotating with $\mb B^{C}(T)$, since $\bm \omega=-\omega \mb u_{X}$ is the rotation rate, the time derivatives in the two frames are related by 
\beq
\dT{\mb S}= \left( \dT{\mb S} \right)_{\mathrm {rot}} +\bm \omega \times \mb S=\left( \dT{\mb S} \right)_{\mathrm {rot}}-\omega \mb u_{X'} \times \mb S. \label{eq:timederiv1}
\eeq 
Then, if we set $\omega-\frac 1 c B^{I}=\omega-\Omega=\Delta \omega$  the spin evolution equation in the rotating frame turns out to be:
\beq
\left( \dT{\mb S} \right)_{\mathrm {rot}}=\left[\Delta \omega \mb u_{X'}+\frac 1 c \mb B^{C}  \right] \times \mb S =\frac 1 c \mb B_{eff} \times \mb S \label{eq:precBeff}
\eeq
In summary, in the rotating frame,  the spinning particle undergoes a precession around the \textit{static} effective gravito-magnetic field  $\displaystyle \mb B_{eff} = c \left[ \Delta \omega \mb u_{X'} +\frac 1 c \mb B^{C} \right]$.
Let us define  $\frac{1}{c}B^{C}=\omega^{*}$; then eq.  (\ref{eq:precBeff}) suggests that if $\Delta \omega \gg \omega^{*}$, the precession is in practice around $\mb u_{X'}$; but if  $\Delta \omega \simeq 0$, i.e.  in \textit{resonance} condition, the spin precession is around  the direction of $\mb B^{C}$, which is in any case in the $YZ$ plane, so the precession may flip the spin completely. This condition is obtained  when the rotation rate of the frame is equal to the frequency of the gravitational wave. In this case, the spin precesses with frequency $\omega^{*}$. Notice that all precessions are referred with respect to the reference spinning particle \cite{biniortolan2017}, at the origin of the Fermi frame so, in any case, we are talking about a \textit{relative precession.}}

The above description is analogous to the classical dynamics of a magnetic moment $\bm \mu$ in a magnetic field $\mb B(t) =\mb B_{0}+\mb B_{1}(t)$  that is the sum of a static field $\mb B_{0}$ and a field $\mb B_{1}(t)$ rotating with frequency $\omega$ in a plane perpendicular to $\mb B_{0}$.  The description of this electromagnetic interaction can be formulated in quantum terms (see e.g. \cite{cohen1991quantum}) as follows. Let us suppose that $|g>$ and $|e>$ are the two eigenvectors, respectively of the ground and excited states,  of the projection of the spinning particle along the $X$ axis;  the eigenvalues of the spin operator $ S_{X}$ are $+\hbar/2$ and $-\hbar/2$, respectively. The Hamiltonian operator is $H=-\bm \mu \cdot \left[\mb B_{0}+\mb B_{1}(t)\right]=-\gamma \mb S \cdot \left[\mb B_{0}+\mb B_{1}(t)\right]$, where $\gamma$ is the gyromagnetic ratio. We set $\omega_{0}=-\gamma B_{0}$, $\omega_{1}=-\gamma B_{1}$. If we suppose that a spin is, at $t=0$, in the ground state $|g>$, the probability of transition to the excited state
 $|e>$ at time $t$ is given by Rabi's formula
\beq
P_{g \rightarrow e}(t)=\frac{(\omega_{1}) ^{2}}{(\omega_{1})^{2}+\left(\omega-\omega_{0}\right)^{2}} \sin^{2} \left (\sqrt{(\omega_{1})^{2}+\left(\omega-\omega_{0}\right)^{2}} \frac t 2 \right) \label{ee:rabi1}
\eeq
We see that for $|\omega-\omega_{0}| \gg |\omega_{1}|$ the probability is almost equal to zero, however when $\omega=\omega_{0}$ we have the magnetic resonance phenomenon, since  $P_{g \rightarrow e}=1$ for $t=\frac{2n+1}{(\omega_{1})} \pi$: in other words, at resonance the oscillation probability does not depend on the rotating magnetic field $\mb B_{1}(t)$ and even a weak field can provoke the flip of the spin direction. We emphasize that in this approach the magnetic filed acts as a purely classical quantity.

According to our approach, in the gravitational wave spacetime there is a gravito-magnetic field $\mb B=\mb B^{I}+\mb B^{C}(T)$.  It is possible to show \cite{Mashhoonspin,Mashhoon:2003ax} that the interaction Hamiltonian of a spin $\mb S$ with the gravito-magnetic field $\mb B$ is $H=\frac 1 c \mb S \cdot \mb B$, and this result can be extended also to the intrinsic spin of particles \cite{hehl1990inertial,ryder1998relativistic}. As a consequence, we may introduce a probability transition for spinning particles in the field of a gravitational wave. Accordingly, we obtain
\beq
P_{g \rightarrow e}(T)=\frac{(\omega^{*}) ^{2}}{(\omega^{*})^{2}+\Delta \omega^{2}} \sin^{2} \left (\sqrt{(\omega^{*})^{2}+\Delta \omega^{2}} \frac T 2 \right) \label{ee:rabi1}
\eeq
Again, at resonance, i.e. when $\Delta \omega=0$, or $\omega=\Omega$, even a weak gravitational field can reverse the direction of the spin: the probability of transition is equal to 1 independently of the strength of the gravito-magnetic field for 
$T=\frac{2n+1}{(\omega^{*})} \pi$.

The resonance condition is obtained by combining the gravito-magnetic field of the wave with a rotation field with the same frequency. Since rotations of the frame cannot be obtained for arbitrary frequencies (it is quite impossible to exceed $10^{3}$ Hz for macroscopic systems), if we are dealing with \textit{charged spinning particles} we may get an equivalent situation by using a true magnetic field, on the basis of Larmor theorem, which  states the equivalence between a system of electric charges in a magnetic field, and the same system rotating with the Larmor frequency. If $B$ denotes the magnitude of the magnetic field, the Larmor frequency for electrons is $\omega_{L}=\frac{\mu_{\mathrm B}}{\hbar } B$, where $\mu_{\mathrm B}$ is the Bohr magneton.  Hence, a true magnetic field  can be used to produce the gravito-magnetic field $\mb B^{I}$.

{This interesting resonance condition between Larmor and gravitational wave frequencies occurring for spinning particles can be easily translated into a resonance of the magnetization of a ferro/-ferrimagnetic material.
In fact, if the sample is a sphere of a cubic crystal magnetized along its symmetry axis, then its magnetization exhibits magnetostatic modes (MSM) with fundamental frequency
$\omega_0(B_0)=\gamma B_0$, with $\gamma = 2 \pi 28$ GHz/T.  The fundamental mode is also known as \textit{Kittel mode} and corresponds to ferromagnetic resonance (FMR) with a uniform magnetization. Clearly this spin-0 mode does not couple to the gravito-magnetic field of the wave.
However, higher MSM modes known as spin waves have
non uniform magnetization. The MSM resonant frequencies can be derived from the solution of the magnetostatic equation \cite{PhysRev.112.59} of a sphere in terms of the associated Legendre Functions $P_n^m(\omega,B_0)$.
In particular, for modes with $n=|m|$ the relation between the resonant frequency of MSM and the external magnetic field is linear and read \cite{gurevich1996magnetization}
\beq
\omega_{m,m,0}=\omega_{H}+\frac{m}{2m+1}\omega_{M} \equiv \gamma H_{e0}+\gamma \frac{4\pi}{3}\frac{m-1}{2m+1}M_{0}, \label{eq:formula1}
\eeq
in terms of the steady field $H_{e0}$ and magnetization $M_{0}$, which is supposed to be along the $X$ axis. Accordingly, for $n=m=2$ the spatial dependence of the magnetization    components turns out to be \cite{gurevich1996magnetization}:
\beq
m_{Y}= Y-i\,Z, \quad m_{Z}=i\, Y+Z. \label{eq:formula2}
\eeq
We notice the quadrupolar-like behaviour for this mode and the correspondence with the  precession determined by the quadrupolar gravito-magnetic field of the wave (\ref{eq:defBxyz12}). A possible approach to the detection of these effects could  be obtained by considering the hybridization of microwave-frequency cavity modes with collective spin excitations, such as
the interaction among the magnetization precession modes in a small magnetically saturated YIG sphere and the MW electromagnetic modes resonating in a RF cavity \cite{PhysRevB.101.014439}.}

\section{Conclusions}\label{sec:dconc}

The construction of the Fermi frame  enables to describe the field of a  gravitational wave in terms of a gravito-electromagnetic analogy; in other words,  the wave field is equivalent to the action of  tidal gravito-electric and a gravito-magnetic fields, that are transverse to the propagation direction and orthogonal to each other.  As for the gravito-magnetic part of wave field, it acts on moving or  spinning  particles; in particular, we have shown that, in analogy with what happens in electromagnetism, a gravitational magnetic resonance phenomenon may appear. Namely, in the Fermi frame, the total gravito-magnetic field is made of a curvature contribution, due to the gravitational wave, and an inertial contribution, due to the rotation rate of the frame. The  gravito-magnetic resonance is produced  when the frame rotates along the direction of propagation of the wave and the rotation rate is equal to the wave frequency. Since the precession frequency is proportional to the square of the wave frequency, high frequency waves (of the order of GHz)  are favoured. If we are dealing with a quantum description of spinning particles, the resonance phenomenon means that the transition probability  reaches the value 1 at suitable times; moreover, this probability does not depend on the gravito-magnetic field, and even a weak field can provoke the flip of the spin direction. However, since it is not possible to have physical rotations for arbitrary frequencies,  we suggested that an equivalent  situation can be obtained by using a true magnetic field, on the basis of the Larmor theorem. Hence, a static magnetic field acting on the probe can mimic the action of a rotating frame. 
Just like in a magnetic resonance phenomenon, it is not the spin of a single particle that can be observed, but that of a great number of identical particles. For instance, the precession induced by  the gravitational wave, can modify the magnetization of a sample. {We suggested that the hybridization of microwave-frequency cavity modes with collective spin excitations could be used to measure these effects.} However, such an analysis is beyond the scope of this paper, whose aim was just to suggest the possibility of considering the phenomenon of gravito-magnetic resonance as the basis of new gravitational waves detection techniques.

\bibliography{SI_draft}

\begin{thebibliography}{30}
\expandafter\ifx\csname natexlab\endcsname\relax\def\natexlab#1{#1}\fi
\expandafter\ifx\csname bibnamefont\endcsname\relax
  \def\bibnamefont#1{#1}\fi
\expandafter\ifx\csname bibfnamefont\endcsname\relax
  \def\bibfnamefont#1{#1}\fi
\expandafter\ifx\csname citenamefont\endcsname\relax
  \def\citenamefont#1{#1}\fi
\expandafter\ifx\csname url\endcsname\relax
  \def\url#1{\texttt{#1}}\fi
\expandafter\ifx\csname urlprefix\endcsname\relax\def\urlprefix{URL }\fi
\providecommand{\bibinfo}[2]{#2}
\providecommand{\eprint}[2][]{\url{#2}}

\bibitem[{\citenamefont{Abbott et~al.}(2016)\citenamefont{Abbott, Abbott,
  Abbott, Abernathy, Acernese, Ackley, Adams, Adams, Addesso, Adhikari
  et~al.}}]{PhysRevLett.116.061102}
\bibinfo{author}{\bibfnamefont{B.~P.} \bibnamefont{Abbott}},
  \bibinfo{author}{\bibfnamefont{R.}~\bibnamefont{Abbott}},
  \bibinfo{author}{\bibfnamefont{T.~D.} \bibnamefont{Abbott}},
  \bibinfo{author}{\bibfnamefont{M.~R.} \bibnamefont{Abernathy}},
  \bibinfo{author}{\bibfnamefont{F.}~\bibnamefont{Acernese}},
  \bibinfo{author}{\bibfnamefont{K.}~\bibnamefont{Ackley}},
  \bibinfo{author}{\bibfnamefont{C.}~\bibnamefont{Adams}},
  \bibinfo{author}{\bibfnamefont{T.}~\bibnamefont{Adams}},
  \bibinfo{author}{\bibfnamefont{P.}~\bibnamefont{Addesso}},
  \bibinfo{author}{\bibfnamefont{R.~X.} \bibnamefont{Adhikari}},
  \bibnamefont{et~al.} (\bibinfo{collaboration}{LIGO Scientific Collaboration
  and Virgo Collaboration}), \bibinfo{journal}{Phys. Rev. Lett.}
  \textbf{\bibinfo{volume}{116}}, \bibinfo{pages}{061102}
  (\bibinfo{year}{2016}),
  \urlprefix\url{https://link.aps.org/doi/10.1103/PhysRevLett.116.061102}.

\bibitem[{\citenamefont{Abbott et~al.}(2017)\citenamefont{Abbott, Abbott,
  Abbott, Acernese, Ackley, Adams, Adams, Addesso, Adhikari, Adya
  et~al.}}]{PhysRevLett.119.161101}
\bibinfo{author}{\bibfnamefont{B.~P.} \bibnamefont{Abbott}},
  \bibinfo{author}{\bibfnamefont{R.}~\bibnamefont{Abbott}},
  \bibinfo{author}{\bibfnamefont{T.~D.} \bibnamefont{Abbott}},
  \bibinfo{author}{\bibfnamefont{F.}~\bibnamefont{Acernese}},
  \bibinfo{author}{\bibfnamefont{K.}~\bibnamefont{Ackley}},
  \bibinfo{author}{\bibfnamefont{C.}~\bibnamefont{Adams}},
  \bibinfo{author}{\bibfnamefont{T.}~\bibnamefont{Adams}},
  \bibinfo{author}{\bibfnamefont{P.}~\bibnamefont{Addesso}},
  \bibinfo{author}{\bibfnamefont{R.~X.} \bibnamefont{Adhikari}},
  \bibinfo{author}{\bibfnamefont{V.~B.} \bibnamefont{Adya}},
  \bibnamefont{et~al.} (\bibinfo{collaboration}{LIGO Scientific Collaboration
  and Virgo Collaboration}), \bibinfo{journal}{Phys. Rev. Lett.}
  \textbf{\bibinfo{volume}{119}}, \bibinfo{pages}{161101}
  (\bibinfo{year}{2017}),
  \urlprefix\url{https://link.aps.org/doi/10.1103/PhysRevLett.119.161101}.

\bibitem[{\citenamefont{{Will}}(2015)}]{2014arXiv1409.7871W}
\bibinfo{author}{\bibfnamefont{C.~M.} \bibnamefont{{Will}}}, in
  \emph{\bibinfo{booktitle}{{General Relativity and Gravitation. A Centennial
  Perspective}}}, edited by
  \bibinfo{editor}{\bibfnamefont{A.}~\bibnamefont{{Ashtekar}}},
  \bibinfo{editor}{\bibfnamefont{B.~K.} \bibnamefont{{Berger}}},
  \bibinfo{editor}{\bibfnamefont{J.}~\bibnamefont{{Isenberg}}},
  \bibnamefont{and}
  \bibinfo{editor}{\bibfnamefont{M.}~\bibnamefont{{MacCallum}}}
  (\bibinfo{publisher}{Cambridge University Press, Cambridge},
  \bibinfo{year}{2015}), pp. \bibinfo{pages}{49--96}.

\bibitem[{\citenamefont{Debono and Smoot}(2016)}]{universe2040023}
\bibinfo{author}{\bibfnamefont{I.}~\bibnamefont{Debono}} \bibnamefont{and}
  \bibinfo{author}{\bibfnamefont{G.~F.} \bibnamefont{Smoot}},
  \bibinfo{journal}{Universe} \textbf{\bibinfo{volume}{2}}
  (\bibinfo{year}{2016}), ISSN \bibinfo{issn}{2218-1997},
  \urlprefix\url{http://www.mdpi.com/2218-1997/2/4/23}.

\bibitem[{\citenamefont{De~Felice and Bini}(2010)}]{de2010classical}
\bibinfo{author}{\bibfnamefont{F.}~\bibnamefont{De~Felice}} \bibnamefont{and}
  \bibinfo{author}{\bibfnamefont{D.}~\bibnamefont{Bini}},
  \emph{\bibinfo{title}{Classical measurements in curved space-times}}
  (\bibinfo{publisher}{Cambridge University Press}, \bibinfo{year}{2010}).

\bibitem[{\citenamefont{Faraoni}(2007)}]{Faraoni:2007gr}
\bibinfo{author}{\bibfnamefont{V.}~\bibnamefont{Faraoni}},
  \bibinfo{journal}{Gen. Rel. Grav.} \textbf{\bibinfo{volume}{39}},
  \bibinfo{pages}{677} (\bibinfo{year}{2007}), \eprint{gr-qc/0702079}.

\bibitem[{\citenamefont{Finn}(2009)}]{Finn:2008np}
\bibinfo{author}{\bibfnamefont{L.~S.} \bibnamefont{Finn}},
  \bibinfo{journal}{Phys. Rev.} \textbf{\bibinfo{volume}{D79}},
  \bibinfo{pages}{022002} (\bibinfo{year}{2009}), \eprint{0810.4529}.

\bibitem[{\citenamefont{Rakhmanov}(2014)}]{Rakhmanov_2014}
\bibinfo{author}{\bibfnamefont{M.}~\bibnamefont{Rakhmanov}},
  \bibinfo{journal}{Classical and Quantum Gravity}
  \textbf{\bibinfo{volume}{31}}, \bibinfo{pages}{085006}
  (\bibinfo{year}{2014}),
  \urlprefix\url{https://doi.org/10.1088%2F0264-9381%2F31%2F8%2F085006}.

\bibitem[{\citenamefont{Fadeev et~al.}(2020)\citenamefont{Fadeev, Wang, Band,
  Budker, Graham, Sushkov, and Kimball}}]{Fadeev:2020gjk}
\bibinfo{author}{\bibfnamefont{P.}~\bibnamefont{Fadeev}},
  \bibinfo{author}{\bibfnamefont{T.}~\bibnamefont{Wang}},
  \bibinfo{author}{\bibfnamefont{Y.}~\bibnamefont{Band}},
  \bibinfo{author}{\bibfnamefont{D.}~\bibnamefont{Budker}},
  \bibinfo{author}{\bibfnamefont{P.~W.} \bibnamefont{Graham}},
  \bibinfo{author}{\bibfnamefont{A.~O.} \bibnamefont{Sushkov}},
  \bibnamefont{and} \bibinfo{author}{\bibfnamefont{D.~F.~J.}
  \bibnamefont{Kimball}} (\bibinfo{year}{2020}), \eprint{2006.09334}.

\bibitem[{\citenamefont{Ruggiero and Tartaglia}(2002)}]{Ruggiero:2002hz}
\bibinfo{author}{\bibfnamefont{M.~L.} \bibnamefont{Ruggiero}} \bibnamefont{and}
  \bibinfo{author}{\bibfnamefont{A.}~\bibnamefont{Tartaglia}},
  \bibinfo{journal}{Nuovo Cim.} \textbf{\bibinfo{volume}{B117}},
  \bibinfo{pages}{743} (\bibinfo{year}{2002}), \eprint{gr-qc/0207065}.

\bibitem[{\citenamefont{Mashhoon}(2003)}]{Mashhoon:2003ax}
\bibinfo{author}{\bibfnamefont{B.}~\bibnamefont{Mashhoon}}
  (\bibinfo{year}{2003}), \eprint{gr-qc/0311030}.

\bibitem[{\citenamefont{Hehl and Ni}(1990)}]{hehl1990inertial}
\bibinfo{author}{\bibfnamefont{F.~W.} \bibnamefont{Hehl}} \bibnamefont{and}
  \bibinfo{author}{\bibfnamefont{W.-T.} \bibnamefont{Ni}},
  \bibinfo{journal}{Physical Review D} \textbf{\bibinfo{volume}{42}},
  \bibinfo{pages}{2045} (\bibinfo{year}{1990}).

\bibitem[{\citenamefont{Tasson}(2012)}]{Tasson:2012nx}
\bibinfo{author}{\bibfnamefont{J.~D.} \bibnamefont{Tasson}},
  \bibinfo{journal}{Phys. Rev. D} \textbf{\bibinfo{volume}{86}},
  \bibinfo{pages}{124021} (\bibinfo{year}{2012}), \eprint{1211.4850}.

\bibitem[{\citenamefont{Hehl et~al.}(1976)\citenamefont{Hehl, Von Der~Heyde,
  Kerlick, and Nester}}]{Hehl:1976kj}
\bibinfo{author}{\bibfnamefont{F.}~\bibnamefont{Hehl}},
  \bibinfo{author}{\bibfnamefont{P.}~\bibnamefont{Von Der~Heyde}},
  \bibinfo{author}{\bibfnamefont{G.}~\bibnamefont{Kerlick}}, \bibnamefont{and}
  \bibinfo{author}{\bibfnamefont{J.}~\bibnamefont{Nester}},
  \bibinfo{journal}{Rev. Mod. Phys.} \textbf{\bibinfo{volume}{48}},
  \bibinfo{pages}{393} (\bibinfo{year}{1976}).

\bibitem[{\citenamefont{Ruggiero and Tartaglia}(2003)}]{Ruggiero:2003tw}
\bibinfo{author}{\bibfnamefont{M.~L.} \bibnamefont{Ruggiero}} \bibnamefont{and}
  \bibinfo{author}{\bibfnamefont{A.}~\bibnamefont{Tartaglia}},
  \bibinfo{journal}{Am. J. Phys.} \textbf{\bibinfo{volume}{71}},
  \bibinfo{pages}{1303} (\bibinfo{year}{2003}), \eprint{gr-qc/0306029}.

\bibitem[{\citenamefont{Capozziello and
  De~Laurentis}(2011)}]{Capozziello:2011et}
\bibinfo{author}{\bibfnamefont{S.}~\bibnamefont{Capozziello}} \bibnamefont{and}
  \bibinfo{author}{\bibfnamefont{M.}~\bibnamefont{De~Laurentis}},
  \bibinfo{journal}{Phys. Rept.} \textbf{\bibinfo{volume}{509}},
  \bibinfo{pages}{167} (\bibinfo{year}{2011}), \eprint{1108.6266}.

\bibitem[{\citenamefont{Safronova et~al.}(2018)\citenamefont{Safronova, Budker,
  DeMille, Kimball, Derevianko, and Clark}}]{Safronova:2017xyt}
\bibinfo{author}{\bibfnamefont{M.}~\bibnamefont{Safronova}},
  \bibinfo{author}{\bibfnamefont{D.}~\bibnamefont{Budker}},
  \bibinfo{author}{\bibfnamefont{D.}~\bibnamefont{DeMille}},
  \bibinfo{author}{\bibfnamefont{D.~F.~J.} \bibnamefont{Kimball}},
  \bibinfo{author}{\bibfnamefont{A.}~\bibnamefont{Derevianko}},
  \bibnamefont{and} \bibinfo{author}{\bibfnamefont{C.}~\bibnamefont{Clark}},
  \bibinfo{journal}{Rev. Mod. Phys.} \textbf{\bibinfo{volume}{90}},
  \bibinfo{pages}{025008} (\bibinfo{year}{2018}), \eprint{1710.01833}.

\bibitem[{\citenamefont{Ruggiero and Ortolan}(2020)}]{Ruggiero_2020}
\bibinfo{author}{\bibfnamefont{M.~L.} \bibnamefont{Ruggiero}} \bibnamefont{and}
  \bibinfo{author}{\bibfnamefont{A.}~\bibnamefont{Ortolan}},
  \bibinfo{journal}{Journal of Physics Communications}
  \textbf{\bibinfo{volume}{4}}, \bibinfo{pages}{055013} (\bibinfo{year}{2020}),
  \urlprefix\url{https://doi.org/10.1088%2F2399-6528%2Fab9320}.

\bibitem[{\citenamefont{Bini et~al.}(2017)\citenamefont{Bini, Geralico, and
  Ortolan}}]{biniortolan2017}
\bibinfo{author}{\bibfnamefont{D.}~\bibnamefont{Bini}},
  \bibinfo{author}{\bibfnamefont{A.}~\bibnamefont{Geralico}}, \bibnamefont{and}
  \bibinfo{author}{\bibfnamefont{A.}~\bibnamefont{Ortolan}},
  \bibinfo{journal}{Phys. Rev. D} \textbf{\bibinfo{volume}{95}},
  \bibinfo{pages}{104044} (\bibinfo{year}{2017}),
  \urlprefix\url{https://link.aps.org/doi/10.1103/PhysRevD.95.104044}.

\bibitem[{\citenamefont{Ito and Soda}(2020)}]{Ito:2020wxi}
\bibinfo{author}{\bibfnamefont{A.}~\bibnamefont{Ito}} \bibnamefont{and}
  \bibinfo{author}{\bibfnamefont{J.}~\bibnamefont{Soda}},
  \bibinfo{journal}{Eur. Phys. J. C} \textbf{\bibinfo{volume}{80}},
  \bibinfo{pages}{545} (\bibinfo{year}{2020}), \eprint{2004.04646}.

\bibitem[{\citenamefont{Misner et~al.}(1973)\citenamefont{Misner, Thorne, and
  Wheeler}}]{MTW}
\bibinfo{author}{\bibfnamefont{C.~W.} \bibnamefont{Misner}},
  \bibinfo{author}{\bibfnamefont{K.~S.} \bibnamefont{Thorne}},
  \bibnamefont{and} \bibinfo{author}{\bibfnamefont{J.~A.}
  \bibnamefont{Wheeler}}, \emph{\bibinfo{title}{Gravitation}}
  (\bibinfo{publisher}{San Francisco: WH Freeman and Co.},
  \bibinfo{year}{1973}).

\bibitem[{\citenamefont{Marzlin}(1994)}]{marzlin}
\bibinfo{author}{\bibfnamefont{K.-P.} \bibnamefont{Marzlin}},
  \bibinfo{journal}{Phys. Rev. D} \textbf{\bibinfo{volume}{50}},
  \bibinfo{pages}{888} (\bibinfo{year}{1994}),
  \urlprefix\url{https://link.aps.org/doi/10.1103/PhysRevD.50.888}.

\bibitem[{\citenamefont{Ni and Zimmermann}(1978)}]{Ni:1978di}
\bibinfo{author}{\bibfnamefont{W.-T.} \bibnamefont{Ni}} \bibnamefont{and}
  \bibinfo{author}{\bibfnamefont{M.}~\bibnamefont{Zimmermann}},
  \bibinfo{journal}{Physical Review D - Particles and Fields}
  \textbf{\bibinfo{volume}{17}}, \bibinfo{pages}{1473} (\bibinfo{year}{1978}).

\bibitem[{\citenamefont{Li and Ni}(1979)}]{Li:1979bz}
\bibinfo{author}{\bibfnamefont{W.-Q.} \bibnamefont{Li}} \bibnamefont{and}
  \bibinfo{author}{\bibfnamefont{W.-T.} \bibnamefont{Ni}},
  \bibinfo{journal}{Journal of Mathematical Physics}
  \textbf{\bibinfo{volume}{20}}, \bibinfo{pages}{1473} (\bibinfo{year}{1979}).

\bibitem[{\citenamefont{Cohen-Tannoudji
  et~al.}(1991)\citenamefont{Cohen-Tannoudji, Diu, and
  Laloe}}]{cohen1991quantum}
\bibinfo{author}{\bibfnamefont{C.}~\bibnamefont{Cohen-Tannoudji}},
  \bibinfo{author}{\bibfnamefont{B.}~\bibnamefont{Diu}}, \bibnamefont{and}
  \bibinfo{author}{\bibfnamefont{F.}~\bibnamefont{Laloe}},
  \emph{\bibinfo{title}{Quantum Mechanics}}, Quantum Mechanics
  (\bibinfo{publisher}{Wiley}, \bibinfo{year}{1991}).

\bibitem[{\citenamefont{Mashhoon}(2000)}]{Mashhoonspin}
\bibinfo{author}{\bibfnamefont{B.}~\bibnamefont{Mashhoon}},
  \bibinfo{journal}{Classical and Quantum Gravity}
  \textbf{\bibinfo{volume}{17}}, \bibinfo{pages}{2399} (\bibinfo{year}{2000}),
  \urlprefix\url{https://doi.org/10.1088%2F0264-9381%2F17%2F12%2F312}.

\bibitem[{\citenamefont{Ryder}(1998)}]{ryder1998relativistic}
\bibinfo{author}{\bibfnamefont{L.}~\bibnamefont{Ryder}},
  \bibinfo{journal}{Journal of Physics A: Mathematical and General}
  \textbf{\bibinfo{volume}{31}}, \bibinfo{pages}{2465} (\bibinfo{year}{1998}).

\bibitem[{\citenamefont{Dillon}(1958)}]{PhysRev.112.59}
\bibinfo{author}{\bibfnamefont{J.~F.} \bibnamefont{Dillon}},
  \bibinfo{journal}{Phys. Rev.} \textbf{\bibinfo{volume}{112}},
  \bibinfo{pages}{59} (\bibinfo{year}{1958}),
  \urlprefix\url{https://link.aps.org/doi/10.1103/PhysRev.112.59}.

\bibitem[{\citenamefont{Gurevich and Melkov}(1996)}]{gurevich1996magnetization}
\bibinfo{author}{\bibfnamefont{A.~G.} \bibnamefont{Gurevich}} \bibnamefont{and}
  \bibinfo{author}{\bibfnamefont{G.~A.} \bibnamefont{Melkov}},
  \emph{\bibinfo{title}{Magnetization oscillations and waves}}
  (\bibinfo{publisher}{CRC press}, \bibinfo{year}{1996}).

\bibitem[{\citenamefont{Leo et~al.}(2020)\citenamefont{Leo, Monteduro, Rizzato,
  Martina, and Maruccio}}]{PhysRevB.101.014439}
\bibinfo{author}{\bibfnamefont{A.}~\bibnamefont{Leo}},
  \bibinfo{author}{\bibfnamefont{A.~G.} \bibnamefont{Monteduro}},
  \bibinfo{author}{\bibfnamefont{S.}~\bibnamefont{Rizzato}},
  \bibinfo{author}{\bibfnamefont{L.}~\bibnamefont{Martina}}, \bibnamefont{and}
  \bibinfo{author}{\bibfnamefont{G.}~\bibnamefont{Maruccio}},
  \bibinfo{journal}{Phys. Rev. B} \textbf{\bibinfo{volume}{101}},
  \bibinfo{pages}{014439} (\bibinfo{year}{2020}),
  \urlprefix\url{https://link.aps.org/doi/10.1103/PhysRevB.101.014439}.

\end{thebibliography}


\begin{thebibliography}{200}


\end{thebibliography}

\end{document}